\pdfoutput=1
\documentclass[aps,prl,reprint,groupedaddress]{revtex4-2}

\usepackage[utf8]{inputenc}
\usepackage[T1]{fontenc}
\usepackage{siunitx}
\usepackage{graphicx}
\graphicspath{{figs/}{figs/setup_optical/}{figs/test_mass_front/}}
\usepackage[hidelinks]{hyperref}
\hypersetup{pdfauthor={Daniel Hartwig, Jan Petermann, Roman Schnabel}, pdftitle={Mechanical parametric feedback-cooling for pendulum-based gravity experiments}}
\usepackage{xcolor}
\usepackage{amsmath}
\usepackage{amssymb}
\usepackage{dcolumn}
\usepackage{cleveref}

\usepackage{pgfplots}
\pgfplotsset{compat=1.15}

\renewcommand{\section}[1]{\paragraph{#1}}

\begin{document}

  %\preprint{}

  %\title{Parametric Feedback Damping of a Pendulum Using Interferometric
  %		Readout}
  %\title{Mechanical Parametric Feedback Damping of a Pendulum for Gravity 
  %Experiments}
  \title{Mechanical parametric feedback-cooling for pendulum-based gravity experiments}


  \author{Daniel Hartwig}
  \author{Jan Petermann}
  \author{Roman Schnabel}
  %\email[]{daniel.hartwig@physnet.uni-hamburg.de}
  %\thanks{}
  \affiliation{Institut für Laserphysik und Zentrum für Optische
  Quantentechnologien, Universität Hamburg}


  \begin{abstract}
Gravitational forces that oscillate at audio-band frequencies are measured with masses suspended as pendulums that have resonance frequencies even lower.
If the pendulum is excited by thermal energy or by seismic motion of the environment, the measurement sensitivity is reduced.
Conventionally, this problem is mitigated by seismic isolation and linear damping, potentially combined with cryogenic cooling.
Here, we propose mechanical parametric cooling of the pendulum motion during the gravitational field measurement.
We report a proof of principle demonstration in the seismic noise dominated regime and achieve a damping factor of the pendulum motion of 5.7.
We find a model system for which mechanical parametric feedback cooling reaches the quantum mechanical regime near the ground state.
More feasible applications we anticipate in gravitational-wave detectors.
\end{abstract}

  %\keywords{}

  \maketitle

  \section{Introduction}

Precision experiments on the gravitational field such as gravitational-wave detectors~\cite{Abbott2016short} or measurements of the gravitational constant~\cite{Parks2010} rely on complex pendulum-like suspensions to minimize non-gravitational coupling to movements in the environment~\cite{Robertson2002}.
The pendulum's resonance frequency needs to be smaller than the measurement frequencies, at which the pendulum mass is then quasi-free in the direction of pendulum motion.
At their resonance frequencies, however, these pendulums are very susceptible to external disturbances.
They excite their oscillatory modes, and due to their low mechanical losses the natural ring-down times are very long.
This problem is conventionally mitigated using active linear feedback damping~\cite{Matichard2015} or passive frictional damping~\cite{Plissi2004}.
With active linear feedback, sensor noise as well as actuator noise is translated into  actuation force on the pendulum, adding broadband noise to the measurement.
On the other hand, passive frictional damping results in a stronger coupling to the thermal bath.
Consequentially, with increasing friction, the thermally excited motion of the test mass at frequencies \emph{above} resonance also increases~\cite{Majorana1997}.
An alternative that evades the drawbacks of both aforementioned damping schemes is parametric damping, which utilizes modulation of oscillation parameters instead of external forces or dissipation to realize the damping effect.

Parametric couplings have proven a useful tool in various areas of research, including gravitational wave detection~\cite{Aguiar1991}.
The center-of-mass motion of a laser-trapped nanoparticle was cooled from room temperature to \SI{\sim50}{mK} using parametric feedback on the laser intensity~\cite{Gieseler2012}.
Using parametric gain or damping on an oscillator coupled to a microwave cavity 
poses an additional possibility for manipulating intracavity fields~\cite{Bothner2020}.
Furthermore, dispersive optomechanical cooling employs an optical cavity of which the resonance frequency is (parametrically) modulated by the position of the mechanical oscillator to be cooled~\cite{Schliesser2006, Schliesser2008, Aspelmeyer2014}.
Average mechanical mode occupation numbers below unity were achieved~\cite{Teufel2011, Chan2011}.
In optics, nonlinear parametric processes enable the reduction of power fluctuations below those of a displaced ground state (coherent state)~\cite{Wu1986, Schnabel2017}.
Parametric amplification was previously used in micro-electromechanical systems to overcome readout noise~\cite{Sharma2012}.

\begin{figure}
	\includegraphics{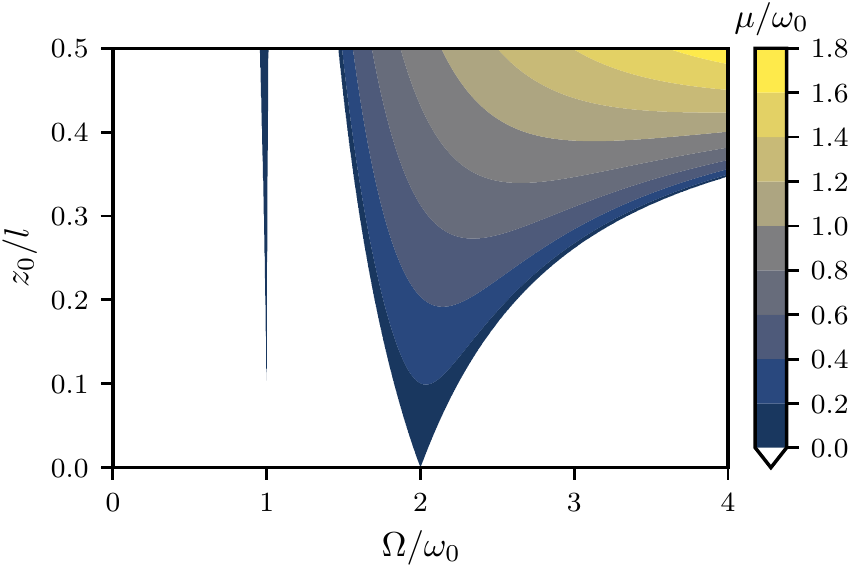}
	\caption{
		Numerically calculated map of the maximal damping rate~$\mu$ of a frictionless parametric oscillator with parametric frequency~$\Omega$ and amplitude~$z_0$.
		The normalizations $\omega_0$ and $l$ are the pendulum's resonance frequency
		and effective pendulum length, respectively.
		For small amplitudes~$z_0 \ll l$, the optimum damping rate is achieved
		for a parametric frequency of $2\omega$. Under this condition the damping
		rate is proportional to $z_0$.}
	\label{fig:stabMap}
\end{figure}

Here, we realize a purely mechanical proof-of-principle experiment, in which we parametrically control the excitation of a pendulum with resonance frequency $\omega_0 = 2\pi \cdot \SI{1.3585}{Hz}$ by vertical actuation of the suspension point at frequency $\Omega = 2\omega_0$.
The equation of motion for the test mass is then given by
\begin{equation}
  \ddot{\phi}(t) + \gamma \dot{\phi}(t) + \frac{1}{l} (g+\ddot{z}(t)) \phi(t) = 0,
  \label{eq:parPendulum}
\end{equation}
where $\phi \ll 1$~is the deflection angle of the pendulum, $\gamma$~is the natural damping rate, defined as the reciprocal of the $1/e$ energy damping time, $l$~is its effective length and $g$~is the gravitational acceleration, which define the resonance frequency~$\omega_0 = \sqrt{g/l}$.
$\ddot{z}(t) = z_0 \Omega^2 \sin \Omega t$ is the vertical acceleration of the suspension point with amplitude~$z_0$, which acts as a parametric drive.
If the suspension moves periodically, energy can be added or removed from the fundamental pendulum oscillation, resulting in an additional damping or heating rate~$\mu$ that adds to the natural damping rate~$\gamma$~\cite{Landau1960}.
\Cref{fig:stabMap} shows the maximal damping~$\mu$ (color coded) versus modulation frequency and amplitude.

Maximal damping requires an optimal phase between the parametric modulation and the oscillation of the pendulum, which is given when the suspension point reaches its maximum vertical speed in direction of the gravitational force each time the pendulum mass crosses its lowest position.
In experiments this phase needs to be servo controlled because it is unstable.
Without this control the phase would converge to the opposite point where instead of damping maximal heating is generated~\cite{Landau1960}.

  \section{Thermal Noise Reduction}

Parametric damping is not only able to compensate transient excitations and seismic noise, but also allows for reducing the root mean square amplitude of the motion below the value of a thermalized state and therefore makes cooling of a single mode possible~\cite{Gieseler2012, Aspelmeyer2014}.

In view of spectral force measurements, the effect of parametric damping has to be described in the frequency picture.
The coupling to a thermal bath can be understood as a random thermal force~$F_\mathrm{th}$ acting on the oscillator.
It has an autocorrelation given by the fluctuation-dissipation-theorem~\cite{Callen1951, Callen1952} as
$\langle F_\mathrm{th}(t) F_\mathrm{th}(t') \rangle = 2 m \gamma k_B T \delta(t-t')$,
where $m$ is the oscillator's effective mass, $\gamma$ is it's natural damping rate, $k_B$ is Boltzmann's constant and $T$ is the temperature.
The single-sided power spectral density of the displacement~$x$ of a purely thermally excited oscillator is then
$S_{xx}(\omega) = |\chi(\omega)|^2 S_{FF,\mathrm{th}}(\omega)$, where $\chi$ is the transfer function of the oscillator and $S_{FF,\mathrm{th}} = 4 m \gamma k_B T$ is the single-sided power spectral density of the thermal force, which is the Fourier-transform of the autocorrelation function $\langle F_\mathrm{th}(t) F_\mathrm{th}(t') \rangle$.
For a harmonic oscillator with the equation of motion $\ddot{x} + \gamma \dot{x} + \omega_0^2 x = \frac{F_\mathrm{th}}{m}$ this evaluates to
\begin{equation}
S_{xx,\mathrm{th}}(\omega) = \frac{4 k_B T}{m} \frac{\gamma}{\left(\omega_0^2 - \omega^2\right)^2 + \gamma^2 \omega^2}.
\label{eq:sxx_harmos}
\end{equation}

Parametric damping by modulation at the frequency~$\Omega = 2\omega_0$ changes the oscillator's displacement spectral density in a way that it modifies the damping rate~($\gamma \rightarrow \gamma + \mu$)~\cite{Landau1960}, yielding
\begin{equation}
	S_{xx,\mathrm{th}}(\omega) = \frac{4 k_B T}{m} \frac{\gamma}{\left(\omega_0^2 - \omega^2\right)^2 + \left(\gamma + \mu\right)^2 \omega^2}.
	\label{eq:sxx_harmos_par}
\end{equation}
The damping rate in the numerator is not modified since it represents the fluctuating thermal force exciting the oscillator, which is not influenced by parametric damping. 
This aspect is key for the cooling capability of parametric damping as we explain below.

\begin{figure}
	\includegraphics{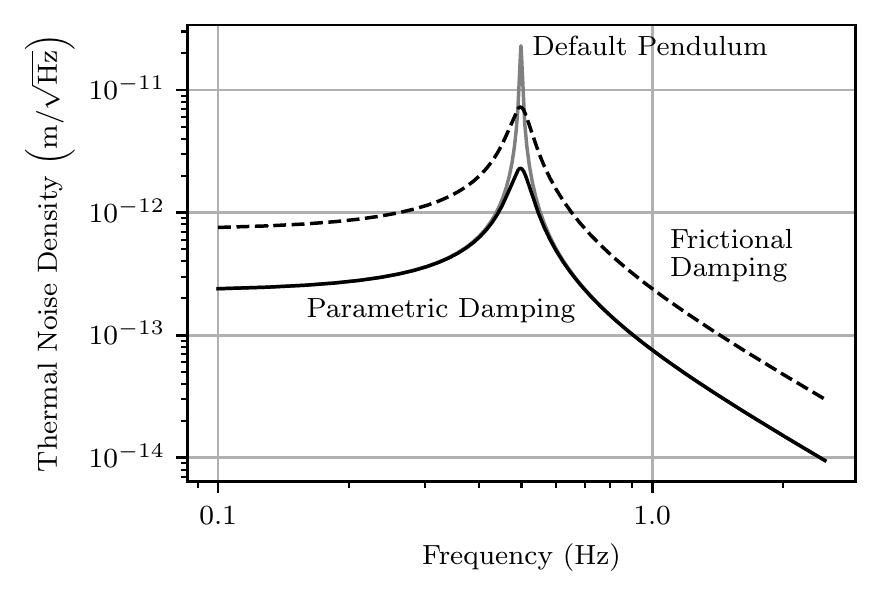}
	\caption{
		Simulated single-sided thermal noise spectral density for a pendulum with a length of \SI{1}{m}, a natural Q-factor of 100 and a mass of \SI{100}{\kilogram} at room temperature (gray).
		The lowest curve shows the same pendulum with parametric damping that increases the damping rate by a factor of 10.
		For comparison, the dashed curve shows the resulting spectral density if the same damping rate was achieved using additional frictional damping.
		In this case, the off-resonance thermal noise is increased, raising the noise floor for measurements at these frequencies.
	}
	\label{fig:psd}
\end{figure}
\Cref{fig:psd} displays the impact of parametric damping on the thermal noise power spectral density of the motion of a thermalized oscillator.
Parametric damping (solid) reduces the on-resonance thermal noise spectral density.
In precision experiments on the gravitational field like interferometric gravitational wave detectors, however, the measurements take place at frequencies \emph{above} the suspension resonances (violin modes of the suspension wires not included)~\cite{Robertson2002}.
Parametric damping on its own won't reduce the thermal noise level there.
This is nevertheless an advantage compared to passive damping, which simply raises the natural damping rate (dashed)~\cite{Gossler2004} since this comes with an increase in off-resonance thermal noise.

\section{Cooling Near The Ground State}
Once a parametrically damped oscillator has reached a stationary state described by the power spectrum in \cref{eq:sxx_harmos_par} we can associate a reduced temperature~$T_r$ based on the mean square position fluctuation~$\left\langle x^2 \right\rangle$ obtained by integrating its power spectral density.
Using the equipartition theorem, $k_B T_r = m \omega_0^2 \left\langle x^2 \right\rangle$, we find
\begin{equation}
\frac{T_r}{T} = \frac{\gamma}{\gamma + \mu}.
\end{equation}
In other words, parametric damping allows the cooling of an oscillator below the temperature of its thermal environment.
To achieve significant temperature reduction relative to the environment, the natural damping rate~$\gamma$ must be small compared to the parametric damping rate~$\mu$.
An oscillator is usually considered to be in its quantum mechanical ground state when the mean phonon occupancy $\langle n \rangle$ is below unity.
Since $\langle n \rangle \hbar \omega_0 = k_B T_r$ we find the requirement for the parametrically induced damping rate to reach the quantum mechanical regime near the ground state
\begin{equation}
\frac{\mu + \gamma}{\gamma} > \frac{k_B T}{\hbar \omega_0}.
\end{equation}
With that we find a model system that would fulfill this condition.
We could imagine a pendulum with a resonance frequency of $\omega_0 = 2\pi \cdot \SI{1}{Hz}$ placed in a dilution refrigerator at a temperature of \SI{2}{\milli\kelvin}~\cite{Bunkov1991}.
In that case the parametric damping rate needs to surpass the natural damping rate by a factor of about \num{4e7}.
If the parametric damping rate corresponds to a Q-factor $Q_\text{par} = \omega_0 / \mu$ as low as 100, which is certainly within reach following the experimental part of this paper, the natural Q-factor $Q_\text{nat} = \omega_0 / \gamma$ needs to have a value around \num{4e9}.
Quality factors in this regime have already been measured on pendulum suspensions in gravitational wave detectors~\cite{Cumming2020}.

  \section{Proof-of-Principle Experiment}
Our setup featured a pendulum which consisted of a cubic test mass suspended by thin wires.
The test mass design can be seen in fig.~\ref{fig:setup_optical}b.
The pendulum assembly was under vacuum.
To decouple the assembly from environmental vibrations, the supporting base plate was suspended by Sorbothane rubber feet inside the vacuum chamber.

The vertical actuation of the suspension point was realized through a commercial piezoelectric actuator inside a flexure housing (Thorlabs APF503) with an actuation range of approximately \SI{0.3}{\milli\meter}.
The connection between the suspension and the test mass was made by two tungsten wire loops that were separated by a few millimeters at the test mass and met at the top where they were clamped.
To minimize the bending losses in the wires, their diameter was chosen as small as possible without risking the wires breaking (\SI{40}{\micro\meter} in our case)~\cite{Cagnoli2000}.

The test mass was a milled aluminum cube with a side length of \SI{33}{mm} and a mass of about \SI{100}{g} (Fig.~\ref{fig:setup_optical}b).
Two opposite sharp edges were aligned horizontally at the same height. This way, the edges provided well-defined break-off points for the wires.
This is expected to minimize the friction of the wire at the contact area~\cite{Cagnoli2000a}.
The two vertically oriented edges featured \ang{45} chamfers of different depths for setting the center of mass slightly below the horizontally oriented edges.
The front surface of the test mass was polished to serve as a mirror.

  \section{Optical Sensing Scheme}

\begin{figure}
\includegraphics{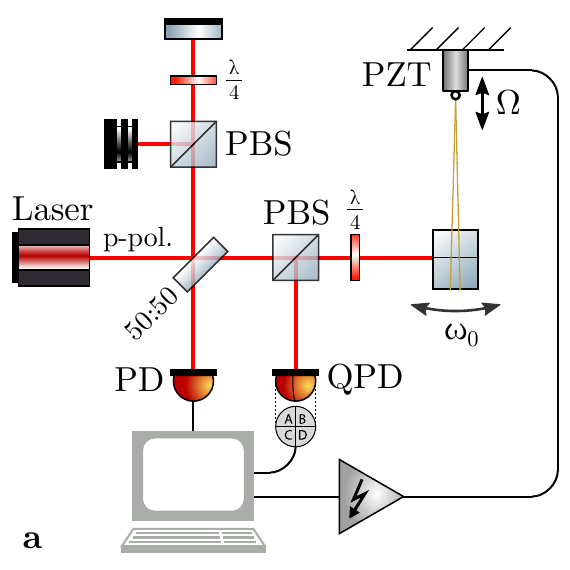}
\includegraphics{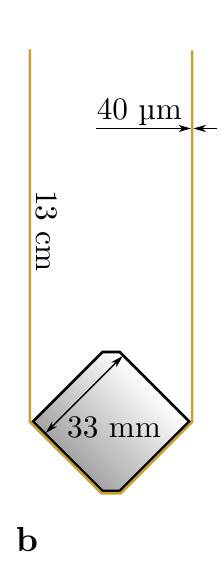}
\caption{
	\textbf{a} Experimental setup.
	The pendulum oscillation at frequency $f = \SI{1.3585}{Hz}$ was parametrically cooled by vertical actuation of the suspension point at frequency $2f$ using using a piezoelectric actuator~(PZT).
	Motion of the pendulum test mass was measured in two ways.
	First, it was interferometrically read out using a photodiode~(PD).
	Second, deflection of the reflected beam was measured using a quadrant photodiode~(QPD) on a part of the reflected beam that was branched off using a polarizing beam splitter (PBS).
	The interferometer offered a higher resolution, but the signal was ambiguous, when the oscillation amplitude was greater than half of a wavelength.
	\textbf{b} Test mass front view.
	The	cubic test mass hanged from a pair of tungsten suspension wires of
	\SI{40}{\micro\meter} thickness. The left and right edges of the cube provided break-off points for the wires.
}
\label{fig:setup_optical}
\end{figure}

The main layout of the optical setup is shown in fig.~\ref{fig:setup_optical}a.
The polished front surface of the pendulum mass reflected a coherent light field of \SI{1550}{nm} wavelength.
The reflected field was analyzed in two ways.
First, a small part of the light returning from the test mass was branched off onto a quadrant segmented photodiode~(QPD).
This allowed the measurement of the beam deflection, which was caused by tilting of the test mass surface in the pendulum mode.
The photocurrents of the upper half of the QPD were subtracted from the ones of the lower half.
This delivered a signal that was roughly proportional to the pendulum's deflection.
Frequency and phase of the pendulum mode were continuously reconstructed from the oscillating signal.
These parameters were then used to adjust the parametric actuation feedback phase to achieve damping.

The amplitude was reconstructed using a Michelson interferometer configuration.
With this, the phase modulation of the reflected field, caused by the back-and-forth motion of the test mass, was converted to an amplitude modulation.
The intensity arriving at the output port was
\begin{equation}
  I(x_m)=\bar{I} \left(1 + V \sin\left(\phi_0 +
  \frac{4\pi x_m}{\lambda}\right)\right),
  \label{eq:interference}
\end{equation}
where $\bar{I}$~is the mean intensity, $V < 1$~is the interference visibility,
$\phi_0$~is a phase offset and $x_m$~is the position of the test mass along the
beam axis.
This nonlinear and ambiguous relationship made it inadequate for phase and frequency reconstruction but gave the ability to measure the oscillation amplitude with more precision than the QPD.
We utilized an algorithm that counted the number of interference fringes crossed per oscillation to reconstruct the amplitude.
Since the pendulum always covered multiple interference fringes during each swing, this method provided sufficiently precise results in our proof-of-principle experiment.

  \section{Achieved Damping Performance}

\begin{figure}
	\includegraphics{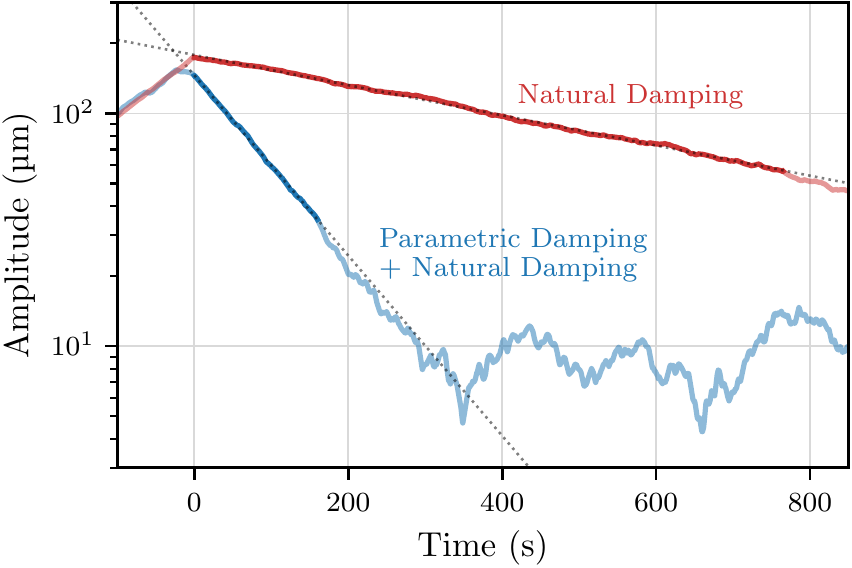}
	\caption{Example ringdown measurements.
	    The pendulum amplitude is recorded over time.
	    In the beginning, the pendulum is parametrically excited, then the
	    parametric damping feedback is either enabled (bottom curve) or left disabled (top curve).
	    The slope during the damping phase can be used to reconstruct the damping rate.
	    This slope is found by fitting a linear function (dashed).
	    With parametric feedback enabled, the damping rate is the sum of parametric and natural damping rates ($\mu + \gamma \approx \SI{17.9e-3}{\per\second}$), without it, it is solely the natural damping rate ($\gamma \approx \SI{3.1e-3}{\per\second}$).
		} 
	\label{fig:ringdownExample}
\end{figure}

To measure the achievable parametric damping rates, the pendulum mode was first parametrically excited to a certain amplitude.
In the second step, the parametric damping feedback loop was enabled and the ringdown of the pendulum amplitude was recorded.
Upon reaching an amplitude of approximately~\SI{10}{\micro\meter}, the QPD was no longer capable of determining the oscillation phase with sufficient precision and the parametric damping was switched off automatically.
An example amplitude measurement is shown in \cref{fig:ringdownExample}.

Once the feedback-loop phase was adjusted for maximal damping, the final measurement was repeated three times.
Four measurements with only the natural damping after excitation served as the reference.
For each measurement, the overall damping rate was determined by fitting a linear function to the curve of the logarithm of the amplitude as seen in \cref{fig:ringdownExample}.
The average natural Q-factor $Q_\text{nat} = \omega_0 / \gamma$ of the pendulum was measured as $2740 \pm 40$. With parametric damping enabled, the Q-factor dropped to $477 \pm 6$.
This implies that the combined parametric and natural damping rate~$\mu+\gamma$ exceeded the natural damping rate~$\gamma$ by a factor of $5.7$.
If the pendulum had an infinitely high natural Q-factor it would have been reduced to approximately 600 by the achieved parametric damping rate.
A larger actuation amplitude would have reduced the Q-factor further.
This would have required a different actuator since we already utilized the maximum actuation range of $\SI{.3}{\milli\meter}$.

  \section{Discussion}

Our experiment shows that parametric damping is a suitable and effective technique to reduce the Q-factor of the pendulum mode of suspended test masses.
According to theory, the Q-factor reduction is independent of the oscillator amplitude range.
Our proof-of-principle experiment was performed at large amplitudes due to poor isolation from seismics.
The lowest Q-factor achieved, however, was not limited by continuous coupling to seismics but by the resolution of the tilt-measurement of the test mass.
For oscillation amplitudes that are large compared to the sensing wavelength, alternative precise interferometric approaches with high dynamic range are available~\cite{Massey1968, Speake2005, Cooper2018}.
For sub-wavelength amplitudes, as one gets in the thermal noise limited regime, a nonpolarizing homodyne interferometer is optimal (as shown in fig.~\ref{fig:setup_optical}a, but without polarizing optics).
Theory predicts that the optimized phase difference between pendulum oscillation and actuation at twice the frequency does not depend on the oscillation amplitude.
After optimization, the actuation is thus time-independent and may translate to a monochromatic feature in the measurement spectrum if the experiment is not ideal and shows some coupling.
A single sharp line is often not an issue in spectrally analyzed data.
In any case, phase and amplitude of actuation is known and can be efficiently subtracted from the data.

  \section{Conclusion}

Vibration isolation systems and vibration mitigation of cryogenic cooling are limited in their performances.
Therefore, high-Q-factor modes often need to be damped.
In our proof-of-principle experiment, we increased the natural damping rate of a pendulum by a factor of $5.7$ using mechanical parametric cooling.
If our pendulum was excited by thermal energy, the factor achieved corresponded to cooling of the pendulum mode from room temperature to about \SI{50}{\kelvin}.
Hypothetically, starting from a Q-factor of \num{4e9}, a cryo-cooled \SI{1}{Hz} pendulum at \SI{2}{\milli\kelvin} could be cooled near its ground state if the pendulum is parametrically damped to a final Q-factor of 100.\\
According to theory, thermal noise in measurements performed with a pendulum being parametrically cooled does not increase at any frequency, which is an advantage over frictional damping.
Furthermore, unlike damping with linear feedback, parametric damping only creates a monochromatic disturbance in the measurement.
Generally, the applicability of mechanical parametric cooling should not be limited to pendulum modes.
Since vertical actuation of the suspension point modulates the stress in suspension fibers, parametric cooling of a selected violin mode should also be possible, as long as excitation of higher order violin modes is not a concern.
In gravitational-wave detectors, the long ring-down times of the suspension violin modes present a challenge to the interferometer control architecture since an excitation of those modes can result in long down times for the entire detector~\cite{Gossler2004}.
To avoid this problem, violin modes were previously reduced in Q-factor by adding friction~\cite{Gossler2004} or by employing feedback control schemes~\cite{Killbourn1999, Dmitriev2009}.
We expect that mechanical parametric damping represents a valuable alternative to friction based damping.

  \section{Acknowledgments}
This research has been supported by the European Research Council (ERC) project “MassQ” (Grant No. 339897) until 2019, and by the Deutsche Forschungsgemeinschaft (DFG, German Research Foundation) under Germany’s Excellence Strategy EXC 2121 “Quantum Universe”—390833306.

  \bibliographystyle{apsrev4-2}
  \bibliography{extra_bib,mendeley}

\begin{thebibliography}{30}%
\makeatletter
\providecommand \@ifxundefined [1]{%
 \@ifx{#1\undefined}
}%
\providecommand \@ifnum [1]{%
 \ifnum #1\expandafter \@firstoftwo
 \else \expandafter \@secondoftwo
 \fi
}%
\providecommand \@ifx [1]{%
 \ifx #1\expandafter \@firstoftwo
 \else \expandafter \@secondoftwo
 \fi
}%
\providecommand \natexlab [1]{#1}%
\providecommand \enquote  [1]{``#1''}%
\providecommand \bibnamefont  [1]{#1}%
\providecommand \bibfnamefont [1]{#1}%
\providecommand \citenamefont [1]{#1}%
\providecommand \href@noop [0]{\@secondoftwo}%
\providecommand \href [0]{\begingroup \@sanitize@url \@href}%
\providecommand \@href[1]{\@@startlink{#1}\@@href}%
\providecommand \@@href[1]{\endgroup#1\@@endlink}%
\providecommand \@sanitize@url [0]{\catcode `\\12\catcode `\$12\catcode
  `\&12\catcode `\#12\catcode `\^12\catcode `\_12\catcode `\%12\relax}%
\providecommand \@@startlink[1]{}%
\providecommand \@@endlink[0]{}%
\providecommand \url  [0]{\begingroup\@sanitize@url \@url }%
\providecommand \@url [1]{\endgroup\@href {#1}{\urlprefix }}%
\providecommand \urlprefix  [0]{URL }%
\providecommand \Eprint [0]{\href }%
\providecommand \doibase [0]{https://doi.org/}%
\providecommand \selectlanguage [0]{\@gobble}%
\providecommand \bibinfo  [0]{\@secondoftwo}%
\providecommand \bibfield  [0]{\@secondoftwo}%
\providecommand \translation [1]{[#1]}%
\providecommand \BibitemOpen [0]{}%
\providecommand \bibitemStop [0]{}%
\providecommand \bibitemNoStop [0]{.\EOS\space}%
\providecommand \EOS [0]{\spacefactor3000\relax}%
\providecommand \BibitemShut  [1]{\csname bibitem#1\endcsname}%
\let\auto@bib@innerbib\@empty
%</preamble>
\bibitem [{\citenamefont {{Abbott, B. P. et al. (LIGO Scientific Collaboration
  and Virgo Collaboration)}}(2016)}]{Abbott2016short}%
  \BibitemOpen
  \bibfield  {author} {\bibinfo {author} {\bibnamefont {{Abbott, B. P. et al.
  (LIGO Scientific Collaboration and Virgo Collaboration)}}},\ }\href
  {https://doi.org/10.1103/PhysRevLett.116.061102} {\bibfield  {journal}
  {\bibinfo  {journal} {Phys. Rev. Lett.}\ }\textbf {\bibinfo {volume} {116}},\
  \bibinfo {pages} {061102} (\bibinfo {year} {2016})},\ \Eprint
  {https://arxiv.org/abs/1602.03837} {arXiv:1602.03837} \BibitemShut {NoStop}%
\bibitem [{\citenamefont {Parks}\ and\ \citenamefont
  {Faller}(2010)}]{Parks2010}%
  \BibitemOpen
  \bibfield  {author} {\bibinfo {author} {\bibfnamefont {H.~V.}\ \bibnamefont
  {Parks}}\ and\ \bibinfo {author} {\bibfnamefont {J.~E.}\ \bibnamefont
  {Faller}},\ }\href {https://doi.org/10.1103/PhysRevLett.105.110801}
  {\bibfield  {journal} {\bibinfo  {journal} {Phys. Rev. Lett.}\ }\textbf
  {\bibinfo {volume} {105}},\ \bibinfo {pages} {110801} (\bibinfo {year}
  {2010})},\ \Eprint {https://arxiv.org/abs/1008.3203} {arXiv:1008.3203}
  \BibitemShut {NoStop}%
\bibitem [{\citenamefont {Robertson}\ \emph {et~al.}(2002)\citenamefont
  {Robertson}, \citenamefont {Cagnoli}, \citenamefont {Crooks}, \citenamefont
  {Elliffe}, \citenamefont {Faller}, \citenamefont {Fritschel}, \citenamefont
  {Go{\ss}ler}, \citenamefont {Grant}, \citenamefont {Heptonstall},
  \citenamefont {Hough}, \citenamefont {L{\"{u}}ck}, \citenamefont {Mittleman},
  \citenamefont {Perreur-Lloyd}, \citenamefont {Plissi}, \citenamefont {Rowan},
  \citenamefont {Shoemaker}, \citenamefont {Sneddon}, \citenamefont {Strain},
  \citenamefont {Torrie}, \citenamefont {Ward},\ and\ \citenamefont
  {Willems}}]{Robertson2002}%
  \BibitemOpen
  \bibfield  {author} {\bibinfo {author} {\bibfnamefont {N.~A.}\ \bibnamefont
  {Robertson}}, \bibinfo {author} {\bibfnamefont {G.}~\bibnamefont {Cagnoli}},
  \bibinfo {author} {\bibfnamefont {D.~R.~M.}\ \bibnamefont {Crooks}}, \bibinfo
  {author} {\bibfnamefont {E.}~\bibnamefont {Elliffe}}, \bibinfo {author}
  {\bibfnamefont {J.~E.}\ \bibnamefont {Faller}}, \bibinfo {author}
  {\bibfnamefont {P.}~\bibnamefont {Fritschel}}, \bibinfo {author}
  {\bibfnamefont {S.}~\bibnamefont {Go{\ss}ler}}, \bibinfo {author}
  {\bibfnamefont {A.}~\bibnamefont {Grant}}, \bibinfo {author} {\bibfnamefont
  {A.}~\bibnamefont {Heptonstall}}, \bibinfo {author} {\bibfnamefont
  {J.}~\bibnamefont {Hough}}, \bibinfo {author} {\bibfnamefont
  {H.}~\bibnamefont {L{\"{u}}ck}}, \bibinfo {author} {\bibfnamefont
  {R.}~\bibnamefont {Mittleman}}, \bibinfo {author} {\bibfnamefont
  {M.}~\bibnamefont {Perreur-Lloyd}}, \bibinfo {author} {\bibfnamefont {M.~V.}\
  \bibnamefont {Plissi}}, \bibinfo {author} {\bibfnamefont {S.}~\bibnamefont
  {Rowan}}, \bibinfo {author} {\bibfnamefont {D.~H.}\ \bibnamefont
  {Shoemaker}}, \bibinfo {author} {\bibfnamefont {P.~H.}\ \bibnamefont
  {Sneddon}}, \bibinfo {author} {\bibfnamefont {K.~A.}\ \bibnamefont {Strain}},
  \bibinfo {author} {\bibfnamefont {C.~I.}\ \bibnamefont {Torrie}}, \bibinfo
  {author} {\bibfnamefont {H.}~\bibnamefont {Ward}},\ and\ \bibinfo {author}
  {\bibfnamefont {P.}~\bibnamefont {Willems}},\ }\href
  {https://doi.org/10.1088/0264-9381/19/15/311} {\bibfield  {journal} {\bibinfo
   {journal} {Class. Quantum Gravity}\ }\textbf {\bibinfo {volume} {19}},\
  \bibinfo {pages} {4043} (\bibinfo {year} {2002})}\BibitemShut {NoStop}%
\bibitem [{\citenamefont {Matichard}\ \emph {et~al.}(2015)\citenamefont
  {Matichard}, \citenamefont {Lantz}, \citenamefont {Mittleman}, \citenamefont
  {Mason}, \citenamefont {Kissel}, \citenamefont {Abbott}, \citenamefont
  {Biscans}, \citenamefont {McIver}, \citenamefont {Abbott}, \citenamefont
  {Abbott}, \citenamefont {Allwine}, \citenamefont {Barnum}, \citenamefont
  {Birch}, \citenamefont {Celerier}, \citenamefont {Clark}, \citenamefont
  {Coyne}, \citenamefont {DeBra}, \citenamefont {DeRosa}, \citenamefont
  {Evans}, \citenamefont {Foley}, \citenamefont {Fritschel}, \citenamefont
  {Giaime}, \citenamefont {Gray}, \citenamefont {Grabeel}, \citenamefont
  {Hanson}, \citenamefont {Hardham}, \citenamefont {Hillard}, \citenamefont
  {Hua}, \citenamefont {Kucharczyk}, \citenamefont {Landry}, \citenamefont {{Le
  Roux}}, \citenamefont {Lhuillier}, \citenamefont {Macleod}, \citenamefont
  {Macinnis}, \citenamefont {Mitchell}, \citenamefont {O'Reilly}, \citenamefont
  {Ottaway}, \citenamefont {Paris}, \citenamefont {Pele}, \citenamefont {Puma},
  \citenamefont {Radkins}, \citenamefont {Ramet}, \citenamefont {Robinson},
  \citenamefont {Ruet}, \citenamefont {Sarin}, \citenamefont {Shoemaker},
  \citenamefont {Stein}, \citenamefont {Thomas}, \citenamefont {Vargas},
  \citenamefont {Venkateswara}, \citenamefont {Warner},\ and\ \citenamefont
  {Wen}}]{Matichard2015}%
  \BibitemOpen
  \bibfield  {author} {\bibinfo {author} {\bibfnamefont {F.}~\bibnamefont
  {Matichard}}, \bibinfo {author} {\bibfnamefont {B.}~\bibnamefont {Lantz}},
  \bibinfo {author} {\bibfnamefont {R.}~\bibnamefont {Mittleman}}, \bibinfo
  {author} {\bibfnamefont {K.}~\bibnamefont {Mason}}, \bibinfo {author}
  {\bibfnamefont {J.}~\bibnamefont {Kissel}}, \bibinfo {author} {\bibfnamefont
  {B.}~\bibnamefont {Abbott}}, \bibinfo {author} {\bibfnamefont
  {S.}~\bibnamefont {Biscans}}, \bibinfo {author} {\bibfnamefont
  {J.}~\bibnamefont {McIver}}, \bibinfo {author} {\bibfnamefont
  {R.}~\bibnamefont {Abbott}}, \bibinfo {author} {\bibfnamefont
  {S.}~\bibnamefont {Abbott}}, \bibinfo {author} {\bibfnamefont
  {E.}~\bibnamefont {Allwine}}, \bibinfo {author} {\bibfnamefont
  {S.}~\bibnamefont {Barnum}}, \bibinfo {author} {\bibfnamefont
  {J.}~\bibnamefont {Birch}}, \bibinfo {author} {\bibfnamefont
  {C.}~\bibnamefont {Celerier}}, \bibinfo {author} {\bibfnamefont
  {D.}~\bibnamefont {Clark}}, \bibinfo {author} {\bibfnamefont
  {D.}~\bibnamefont {Coyne}}, \bibinfo {author} {\bibfnamefont
  {D.}~\bibnamefont {DeBra}}, \bibinfo {author} {\bibfnamefont
  {R.}~\bibnamefont {DeRosa}}, \bibinfo {author} {\bibfnamefont
  {M.}~\bibnamefont {Evans}}, \bibinfo {author} {\bibfnamefont
  {S.}~\bibnamefont {Foley}}, \bibinfo {author} {\bibfnamefont
  {P.}~\bibnamefont {Fritschel}}, \bibinfo {author} {\bibfnamefont {J.~A.}\
  \bibnamefont {Giaime}}, \bibinfo {author} {\bibfnamefont {C.}~\bibnamefont
  {Gray}}, \bibinfo {author} {\bibfnamefont {G.}~\bibnamefont {Grabeel}},
  \bibinfo {author} {\bibfnamefont {J.}~\bibnamefont {Hanson}}, \bibinfo
  {author} {\bibfnamefont {C.}~\bibnamefont {Hardham}}, \bibinfo {author}
  {\bibfnamefont {M.}~\bibnamefont {Hillard}}, \bibinfo {author} {\bibfnamefont
  {W.}~\bibnamefont {Hua}}, \bibinfo {author} {\bibfnamefont {C.}~\bibnamefont
  {Kucharczyk}}, \bibinfo {author} {\bibfnamefont {M.}~\bibnamefont {Landry}},
  \bibinfo {author} {\bibfnamefont {A.}~\bibnamefont {{Le Roux}}}, \bibinfo
  {author} {\bibfnamefont {V.}~\bibnamefont {Lhuillier}}, \bibinfo {author}
  {\bibfnamefont {D.}~\bibnamefont {Macleod}}, \bibinfo {author} {\bibfnamefont
  {M.}~\bibnamefont {Macinnis}}, \bibinfo {author} {\bibfnamefont
  {R.}~\bibnamefont {Mitchell}}, \bibinfo {author} {\bibfnamefont
  {B.}~\bibnamefont {O'Reilly}}, \bibinfo {author} {\bibfnamefont
  {D.}~\bibnamefont {Ottaway}}, \bibinfo {author} {\bibfnamefont
  {H.}~\bibnamefont {Paris}}, \bibinfo {author} {\bibfnamefont
  {A.}~\bibnamefont {Pele}}, \bibinfo {author} {\bibfnamefont {M.}~\bibnamefont
  {Puma}}, \bibinfo {author} {\bibfnamefont {H.}~\bibnamefont {Radkins}},
  \bibinfo {author} {\bibfnamefont {C.}~\bibnamefont {Ramet}}, \bibinfo
  {author} {\bibfnamefont {M.}~\bibnamefont {Robinson}}, \bibinfo {author}
  {\bibfnamefont {L.}~\bibnamefont {Ruet}}, \bibinfo {author} {\bibfnamefont
  {P.}~\bibnamefont {Sarin}}, \bibinfo {author} {\bibfnamefont
  {D.}~\bibnamefont {Shoemaker}}, \bibinfo {author} {\bibfnamefont
  {A.}~\bibnamefont {Stein}}, \bibinfo {author} {\bibfnamefont
  {J.}~\bibnamefont {Thomas}}, \bibinfo {author} {\bibfnamefont
  {M.}~\bibnamefont {Vargas}}, \bibinfo {author} {\bibfnamefont
  {K.}~\bibnamefont {Venkateswara}}, \bibinfo {author} {\bibfnamefont
  {J.}~\bibnamefont {Warner}},\ and\ \bibinfo {author} {\bibfnamefont
  {S.}~\bibnamefont {Wen}},\ }\href
  {https://doi.org/10.1088/0264-9381/32/18/185003} {\bibfield  {journal}
  {\bibinfo  {journal} {Class. Quantum Gravity}\ }\textbf {\bibinfo {volume}
  {32}},\ \bibinfo {pages} {185003} (\bibinfo {year} {2015})},\ \Eprint
  {https://arxiv.org/abs/1502.06300} {arXiv:1502.06300} \BibitemShut {NoStop}%
\bibitem [{\citenamefont {Plissi}\ \emph {et~al.}(2004)\citenamefont {Plissi},
  \citenamefont {Torrie}, \citenamefont {Barton}, \citenamefont {Robertson},
  \citenamefont {Grant}, \citenamefont {Cantley}, \citenamefont {Strain},
  \citenamefont {Willems}, \citenamefont {Romie}, \citenamefont {Skeldon},
  \citenamefont {Perreur-Lloyd}, \citenamefont {Jones},\ and\ \citenamefont
  {Hough}}]{Plissi2004}%
  \BibitemOpen
  \bibfield  {author} {\bibinfo {author} {\bibfnamefont {M.~V.}\ \bibnamefont
  {Plissi}}, \bibinfo {author} {\bibfnamefont {C.~I.}\ \bibnamefont {Torrie}},
  \bibinfo {author} {\bibfnamefont {M.}~\bibnamefont {Barton}}, \bibinfo
  {author} {\bibfnamefont {N.~A.}\ \bibnamefont {Robertson}}, \bibinfo {author}
  {\bibfnamefont {A.}~\bibnamefont {Grant}}, \bibinfo {author} {\bibfnamefont
  {C.~A.}\ \bibnamefont {Cantley}}, \bibinfo {author} {\bibfnamefont {K.~A.}\
  \bibnamefont {Strain}}, \bibinfo {author} {\bibfnamefont {P.~A.}\
  \bibnamefont {Willems}}, \bibinfo {author} {\bibfnamefont {J.~H.}\
  \bibnamefont {Romie}}, \bibinfo {author} {\bibfnamefont {K.~D.}\ \bibnamefont
  {Skeldon}}, \bibinfo {author} {\bibfnamefont {M.~M.}\ \bibnamefont
  {Perreur-Lloyd}}, \bibinfo {author} {\bibfnamefont {R.~A.}\ \bibnamefont
  {Jones}},\ and\ \bibinfo {author} {\bibfnamefont {J.}~\bibnamefont {Hough}},\
  }\href {https://doi.org/10.1063/1.1795192} {\bibfield  {journal} {\bibinfo
  {journal} {Rev. Sci. Instrum.}\ }\textbf {\bibinfo {volume} {75}},\ \bibinfo
  {pages} {4516} (\bibinfo {year} {2004})}\BibitemShut {NoStop}%
\bibitem [{\citenamefont {Majorana}\ and\ \citenamefont
  {Ogawa}(1997)}]{Majorana1997}%
  \BibitemOpen
  \bibfield  {author} {\bibinfo {author} {\bibfnamefont {E.}~\bibnamefont
  {Majorana}}\ and\ \bibinfo {author} {\bibfnamefont {Y.}~\bibnamefont
  {Ogawa}},\ }\href {https://doi.org/10.1016/S0375-9601(97)00458-1} {\bibfield
  {journal} {\bibinfo  {journal} {Phys. Lett. A}\ }\textbf {\bibinfo {volume}
  {233}},\ \bibinfo {pages} {162} (\bibinfo {year} {1997})}\BibitemShut
  {NoStop}%
\bibitem [{\citenamefont {Aguiar}\ \emph {et~al.}(1991)\citenamefont {Aguiar},
  \citenamefont {Johnson},\ and\ \citenamefont {Hamilton}}]{Aguiar1991}%
  \BibitemOpen
  \bibfield  {author} {\bibinfo {author} {\bibfnamefont {O.~D.}\ \bibnamefont
  {Aguiar}}, \bibinfo {author} {\bibfnamefont {W.~W.}\ \bibnamefont
  {Johnson}},\ and\ \bibinfo {author} {\bibfnamefont {W.~O.}\ \bibnamefont
  {Hamilton}},\ }\href {https://doi.org/10.1063/1.1142226} {\bibfield
  {journal} {\bibinfo  {journal} {Rev. Sci. Instrum.}\ }\textbf {\bibinfo
  {volume} {62}},\ \bibinfo {pages} {2523} (\bibinfo {year}
  {1991})}\BibitemShut {NoStop}%
\bibitem [{\citenamefont {Gieseler}\ \emph {et~al.}(2012)\citenamefont
  {Gieseler}, \citenamefont {Deutsch}, \citenamefont {Quidant},\ and\
  \citenamefont {Novotny}}]{Gieseler2012}%
  \BibitemOpen
  \bibfield  {author} {\bibinfo {author} {\bibfnamefont {J.}~\bibnamefont
  {Gieseler}}, \bibinfo {author} {\bibfnamefont {B.}~\bibnamefont {Deutsch}},
  \bibinfo {author} {\bibfnamefont {R.}~\bibnamefont {Quidant}},\ and\ \bibinfo
  {author} {\bibfnamefont {L.}~\bibnamefont {Novotny}},\ }\href
  {https://doi.org/10.1103/PhysRevLett.109.103603} {\bibfield  {journal}
  {\bibinfo  {journal} {Phys. Rev. Lett.}\ }\textbf {\bibinfo {volume} {109}},\
  \bibinfo {pages} {103603} (\bibinfo {year} {2012})},\ \Eprint
  {https://arxiv.org/abs/1202.6435} {arXiv:1202.6435} \BibitemShut {NoStop}%
\bibitem [{\citenamefont {Bothner}\ \emph {et~al.}(2020)\citenamefont
  {Bothner}, \citenamefont {Yanai}, \citenamefont {Iniguez-Rabago},
  \citenamefont {Yuan}, \citenamefont {Blanter},\ and\ \citenamefont
  {Steele}}]{Bothner2020}%
  \BibitemOpen
  \bibfield  {author} {\bibinfo {author} {\bibfnamefont {D.}~\bibnamefont
  {Bothner}}, \bibinfo {author} {\bibfnamefont {S.}~\bibnamefont {Yanai}},
  \bibinfo {author} {\bibfnamefont {A.}~\bibnamefont {Iniguez-Rabago}},
  \bibinfo {author} {\bibfnamefont {M.}~\bibnamefont {Yuan}}, \bibinfo {author}
  {\bibfnamefont {Y.~M.}\ \bibnamefont {Blanter}},\ and\ \bibinfo {author}
  {\bibfnamefont {G.~A.}\ \bibnamefont {Steele}},\ }\href
  {https://doi.org/10.1038/s41467-020-15389-4} {\bibfield  {journal} {\bibinfo
  {journal} {Nat. Commun.}\ }\textbf {\bibinfo {volume} {11}},\ \bibinfo
  {pages} {1589} (\bibinfo {year} {2020})},\ \Eprint
  {https://arxiv.org/abs/1908.08496} {arXiv:1908.08496} \BibitemShut {NoStop}%
\bibitem [{\citenamefont {Schliesser}\ \emph {et~al.}(2006)\citenamefont
  {Schliesser}, \citenamefont {Del'Haye}, \citenamefont {Nooshi}, \citenamefont
  {Vahala},\ and\ \citenamefont {Kippenberg}}]{Schliesser2006}%
  \BibitemOpen
  \bibfield  {author} {\bibinfo {author} {\bibfnamefont {A.}~\bibnamefont
  {Schliesser}}, \bibinfo {author} {\bibfnamefont {P.}~\bibnamefont
  {Del'Haye}}, \bibinfo {author} {\bibfnamefont {N.}~\bibnamefont {Nooshi}},
  \bibinfo {author} {\bibfnamefont {K.~J.}\ \bibnamefont {Vahala}},\ and\
  \bibinfo {author} {\bibfnamefont {T.~J.}\ \bibnamefont {Kippenberg}},\ }\href
  {https://doi.org/10.1103/PhysRevLett.97.243905} {\bibfield  {journal}
  {\bibinfo  {journal} {Phys. Rev. Lett.}\ }\textbf {\bibinfo {volume} {97}},\
  \bibinfo {pages} {243905} (\bibinfo {year} {2006})},\ \Eprint
  {https://arxiv.org/abs/0611235} {arXiv:0611235 [physics]} \BibitemShut
  {NoStop}%
\bibitem [{\citenamefont {Schliesser}\ \emph {et~al.}(2008)\citenamefont
  {Schliesser}, \citenamefont {Rivi{\`{e}}re}, \citenamefont {Anetsberger},
  \citenamefont {Arcizet},\ and\ \citenamefont {Kippenberg}}]{Schliesser2008}%
  \BibitemOpen
  \bibfield  {author} {\bibinfo {author} {\bibfnamefont {A.}~\bibnamefont
  {Schliesser}}, \bibinfo {author} {\bibfnamefont {R.}~\bibnamefont
  {Rivi{\`{e}}re}}, \bibinfo {author} {\bibfnamefont {G.}~\bibnamefont
  {Anetsberger}}, \bibinfo {author} {\bibfnamefont {O.}~\bibnamefont
  {Arcizet}},\ and\ \bibinfo {author} {\bibfnamefont {T.~J.}\ \bibnamefont
  {Kippenberg}},\ }\href {https://doi.org/10.1038/nphys939} {\bibfield
  {journal} {\bibinfo  {journal} {Nat. Phys.}\ }\textbf {\bibinfo {volume}
  {4}},\ \bibinfo {pages} {415} (\bibinfo {year} {2008})},\ \Eprint
  {https://arxiv.org/abs/0709.4036} {arXiv:0709.4036} \BibitemShut {NoStop}%
\bibitem [{\citenamefont {Aspelmeyer}\ \emph {et~al.}(2014)\citenamefont
  {Aspelmeyer}, \citenamefont {Kippenberg},\ and\ \citenamefont
  {Marquardt}}]{Aspelmeyer2014}%
  \BibitemOpen
  \bibfield  {author} {\bibinfo {author} {\bibfnamefont {M.}~\bibnamefont
  {Aspelmeyer}}, \bibinfo {author} {\bibfnamefont {T.~J.}\ \bibnamefont
  {Kippenberg}},\ and\ \bibinfo {author} {\bibfnamefont {F.}~\bibnamefont
  {Marquardt}},\ }\href {https://doi.org/10.1103/RevModPhys.86.1391} {\bibfield
   {journal} {\bibinfo  {journal} {Rev. Mod. Phys.}\ }\textbf {\bibinfo
  {volume} {86}},\ \bibinfo {pages} {1391} (\bibinfo {year} {2014})},\ \Eprint
  {https://arxiv.org/abs/1303.0733} {arXiv:1303.0733} \BibitemShut {NoStop}%
\bibitem [{\citenamefont {Teufel}\ \emph {et~al.}(2011)\citenamefont {Teufel},
  \citenamefont {Donner}, \citenamefont {Li}, \citenamefont {Harlow},
  \citenamefont {Allman}, \citenamefont {Cicak}, \citenamefont {Sirois},
  \citenamefont {Whittaker}, \citenamefont {Lehnert},\ and\ \citenamefont
  {Simmonds}}]{Teufel2011}%
  \BibitemOpen
  \bibfield  {author} {\bibinfo {author} {\bibfnamefont {J.~D.}\ \bibnamefont
  {Teufel}}, \bibinfo {author} {\bibfnamefont {T.}~\bibnamefont {Donner}},
  \bibinfo {author} {\bibfnamefont {D.}~\bibnamefont {Li}}, \bibinfo {author}
  {\bibfnamefont {J.~W.}\ \bibnamefont {Harlow}}, \bibinfo {author}
  {\bibfnamefont {M.~S.}\ \bibnamefont {Allman}}, \bibinfo {author}
  {\bibfnamefont {K.}~\bibnamefont {Cicak}}, \bibinfo {author} {\bibfnamefont
  {A.~J.}\ \bibnamefont {Sirois}}, \bibinfo {author} {\bibfnamefont {J.~D.}\
  \bibnamefont {Whittaker}}, \bibinfo {author} {\bibfnamefont {K.~W.}\
  \bibnamefont {Lehnert}},\ and\ \bibinfo {author} {\bibfnamefont {R.~W.}\
  \bibnamefont {Simmonds}},\ }\href {https://doi.org/10.1038/nature10261}
  {\bibfield  {journal} {\bibinfo  {journal} {Nature}\ }\textbf {\bibinfo
  {volume} {475}},\ \bibinfo {pages} {359} (\bibinfo {year} {2011})},\ \Eprint
  {https://arxiv.org/abs/1103.2144} {arXiv:1103.2144} \BibitemShut {NoStop}%
\bibitem [{\citenamefont {Chan}\ \emph {et~al.}(2011)\citenamefont {Chan},
  \citenamefont {Alegre}, \citenamefont {Safavi-Naeini}, \citenamefont {Hill},
  \citenamefont {Krause}, \citenamefont {Gr{\"{o}}blacher}, \citenamefont
  {Aspelmeyer},\ and\ \citenamefont {Painter}}]{Chan2011}%
  \BibitemOpen
  \bibfield  {author} {\bibinfo {author} {\bibfnamefont {J.}~\bibnamefont
  {Chan}}, \bibinfo {author} {\bibfnamefont {T.~P.~M.}\ \bibnamefont {Alegre}},
  \bibinfo {author} {\bibfnamefont {A.~H.}\ \bibnamefont {Safavi-Naeini}},
  \bibinfo {author} {\bibfnamefont {J.~T.}\ \bibnamefont {Hill}}, \bibinfo
  {author} {\bibfnamefont {A.}~\bibnamefont {Krause}}, \bibinfo {author}
  {\bibfnamefont {S.}~\bibnamefont {Gr{\"{o}}blacher}}, \bibinfo {author}
  {\bibfnamefont {M.}~\bibnamefont {Aspelmeyer}},\ and\ \bibinfo {author}
  {\bibfnamefont {O.}~\bibnamefont {Painter}},\ }\href
  {https://doi.org/10.1038/nature10461} {\bibfield  {journal} {\bibinfo
  {journal} {Nature}\ }\textbf {\bibinfo {volume} {478}},\ \bibinfo {pages}
  {89} (\bibinfo {year} {2011})},\ \Eprint {https://arxiv.org/abs/1106.3614}
  {arXiv:1106.3614} \BibitemShut {NoStop}%
\bibitem [{\citenamefont {Wu}\ \emph {et~al.}(1986)\citenamefont {Wu},
  \citenamefont {Kimble}, \citenamefont {Hall},\ and\ \citenamefont
  {Wu}}]{Wu1986}%
  \BibitemOpen
  \bibfield  {author} {\bibinfo {author} {\bibfnamefont {L.-A.}\ \bibnamefont
  {Wu}}, \bibinfo {author} {\bibfnamefont {H.~J.}\ \bibnamefont {Kimble}},
  \bibinfo {author} {\bibfnamefont {J.~L.}\ \bibnamefont {Hall}},\ and\
  \bibinfo {author} {\bibfnamefont {H.}~\bibnamefont {Wu}},\ }\href
  {https://doi.org/10.1103/PhysRevLett.57.2520} {\bibfield  {journal} {\bibinfo
   {journal} {Phys. Rev. Lett.}\ }\textbf {\bibinfo {volume} {57}},\ \bibinfo
  {pages} {2520} (\bibinfo {year} {1986})}\BibitemShut {NoStop}%
\bibitem [{\citenamefont {Schnabel}(2017)}]{Schnabel2017}%
  \BibitemOpen
  \bibfield  {author} {\bibinfo {author} {\bibfnamefont {R.}~\bibnamefont
  {Schnabel}},\ }\href {https://doi.org/10.1016/j.physrep.2017.04.001}
  {\bibfield  {journal} {\bibinfo  {journal} {Phys. Rep.}\ }\textbf {\bibinfo
  {volume} {684}},\ \bibinfo {pages} {1} (\bibinfo {year} {2017})},\ \Eprint
  {https://arxiv.org/abs/1611.03986} {arXiv:1611.03986} \BibitemShut {NoStop}%
\bibitem [{\citenamefont {Sharma}\ \emph {et~al.}(2012)\citenamefont {Sharma},
  \citenamefont {Sarraf}, \citenamefont {Baskaran},\ and\ \citenamefont
  {Cretu}}]{Sharma2012}%
  \BibitemOpen
  \bibfield  {author} {\bibinfo {author} {\bibfnamefont {M.}~\bibnamefont
  {Sharma}}, \bibinfo {author} {\bibfnamefont {E.~H.}\ \bibnamefont {Sarraf}},
  \bibinfo {author} {\bibfnamefont {R.}~\bibnamefont {Baskaran}},\ and\
  \bibinfo {author} {\bibfnamefont {E.}~\bibnamefont {Cretu}},\ }\href
  {https://doi.org/10.1016/j.sna.2011.08.009} {\bibfield  {journal} {\bibinfo
  {journal} {Sensors Actuators A Phys.}\ }\textbf {\bibinfo {volume} {177}},\
  \bibinfo {pages} {79} (\bibinfo {year} {2012})}\BibitemShut {NoStop}%
\bibitem [{\citenamefont {Landau}\ and\ \citenamefont
  {Lifshitz}(1960)}]{Landau1960}%
  \BibitemOpen
  \bibfield  {author} {\bibinfo {author} {\bibfnamefont {L.~D.}\ \bibnamefont
  {Landau}}\ and\ \bibinfo {author} {\bibfnamefont {E.~M.}\ \bibnamefont
  {Lifshitz}},\ }\href {https://archive.org/details/Mechanics{\_}541} {\emph
  {\bibinfo {title} {{Mechanics}}}},\ \bibinfo {edition} {2nd}\ ed.\ (\bibinfo
  {publisher} {Pergamon Press},\ \bibinfo {address} {Oxford},\ \bibinfo {year}
  {1960})\BibitemShut {NoStop}%
\bibitem [{\citenamefont {Callen}\ and\ \citenamefont
  {Welton}(1951)}]{Callen1951}%
  \BibitemOpen
  \bibfield  {author} {\bibinfo {author} {\bibfnamefont {H.~B.}\ \bibnamefont
  {Callen}}\ and\ \bibinfo {author} {\bibfnamefont {T.~A.}\ \bibnamefont
  {Welton}},\ }\href {https://doi.org/10.1103/PhysRev.83.34} {\bibfield
  {journal} {\bibinfo  {journal} {Phys. Rev.}\ }\textbf {\bibinfo {volume}
  {83}},\ \bibinfo {pages} {34} (\bibinfo {year} {1951})}\BibitemShut {NoStop}%
\bibitem [{\citenamefont {Callen}\ and\ \citenamefont
  {Greene}(1952)}]{Callen1952}%
  \BibitemOpen
  \bibfield  {author} {\bibinfo {author} {\bibfnamefont {H.~B.}\ \bibnamefont
  {Callen}}\ and\ \bibinfo {author} {\bibfnamefont {R.~F.}\ \bibnamefont
  {Greene}},\ }\href {https://doi.org/10.1103/PhysRev.86.702} {\bibfield
  {journal} {\bibinfo  {journal} {Phys. Rev.}\ }\textbf {\bibinfo {volume}
  {86}},\ \bibinfo {pages} {702} (\bibinfo {year} {1952})}\BibitemShut
  {NoStop}%
\bibitem [{\citenamefont {Go{\ss}ler}\ \emph {et~al.}(2004)\citenamefont
  {Go{\ss}ler}, \citenamefont {Cagnoli}, \citenamefont {Crooks}, \citenamefont
  {L{\"{u}}ck}, \citenamefont {Rowan}, \citenamefont {Smith}, \citenamefont
  {Strain}, \citenamefont {Hough},\ and\ \citenamefont
  {Danzmann}}]{Gossler2004}%
  \BibitemOpen
  \bibfield  {author} {\bibinfo {author} {\bibfnamefont {S.}~\bibnamefont
  {Go{\ss}ler}}, \bibinfo {author} {\bibfnamefont {G.}~\bibnamefont {Cagnoli}},
  \bibinfo {author} {\bibfnamefont {D.~R.~M.}\ \bibnamefont {Crooks}}, \bibinfo
  {author} {\bibfnamefont {H.}~\bibnamefont {L{\"{u}}ck}}, \bibinfo {author}
  {\bibfnamefont {S.}~\bibnamefont {Rowan}}, \bibinfo {author} {\bibfnamefont
  {J.~R.}\ \bibnamefont {Smith}}, \bibinfo {author} {\bibfnamefont {K.~A.}\
  \bibnamefont {Strain}}, \bibinfo {author} {\bibfnamefont {J.}~\bibnamefont
  {Hough}},\ and\ \bibinfo {author} {\bibfnamefont {K.}~\bibnamefont
  {Danzmann}},\ }\href {https://doi.org/10.1088/0264-9381/21/5/082} {\bibfield
  {journal} {\bibinfo  {journal} {Class. Quantum Gravity}\ }\textbf {\bibinfo
  {volume} {21}},\ \bibinfo {pages} {S923} (\bibinfo {year}
  {2004})}\BibitemShut {NoStop}%
\bibitem [{\citenamefont {Bunkov}\ \emph {et~al.}(1991)\citenamefont {Bunkov},
  \citenamefont {Gu{\'{e}}nault}, \citenamefont {Hayward}, \citenamefont
  {Jackson}, \citenamefont {Kennedy}, \citenamefont {Nichols}, \citenamefont
  {Miller}, \citenamefont {Pickett},\ and\ \citenamefont {Ward}}]{Bunkov1991}%
  \BibitemOpen
  \bibfield  {author} {\bibinfo {author} {\bibfnamefont {Y.~M.}\ \bibnamefont
  {Bunkov}}, \bibinfo {author} {\bibfnamefont {A.~M.}\ \bibnamefont
  {Gu{\'{e}}nault}}, \bibinfo {author} {\bibfnamefont {D.~J.}\ \bibnamefont
  {Hayward}}, \bibinfo {author} {\bibfnamefont {D.~A.}\ \bibnamefont
  {Jackson}}, \bibinfo {author} {\bibfnamefont {C.~J.}\ \bibnamefont
  {Kennedy}}, \bibinfo {author} {\bibfnamefont {T.~R.}\ \bibnamefont
  {Nichols}}, \bibinfo {author} {\bibfnamefont {I.~E.}\ \bibnamefont {Miller}},
  \bibinfo {author} {\bibfnamefont {G.~R.}\ \bibnamefont {Pickett}},\ and\
  \bibinfo {author} {\bibfnamefont {M.~G.}\ \bibnamefont {Ward}},\ }\href
  {https://doi.org/10.1007/BF00683626} {\bibfield  {journal} {\bibinfo
  {journal} {J. Low Temp. Phys.}\ }\textbf {\bibinfo {volume} {83}},\ \bibinfo
  {pages} {257} (\bibinfo {year} {1991})}\BibitemShut {NoStop}%
\bibitem [{\citenamefont {Cumming}\ \emph {et~al.}(2020)\citenamefont
  {Cumming}, \citenamefont {Sorazu}, \citenamefont {Daw}, \citenamefont
  {Hammond}, \citenamefont {Hough}, \citenamefont {Jones}, \citenamefont
  {Martin}, \citenamefont {Rowan}, \citenamefont {Strain},\ and\ \citenamefont
  {Williams}}]{Cumming2020}%
  \BibitemOpen
  \bibfield  {author} {\bibinfo {author} {\bibfnamefont {A.~V.}\ \bibnamefont
  {Cumming}}, \bibinfo {author} {\bibfnamefont {B.}~\bibnamefont {Sorazu}},
  \bibinfo {author} {\bibfnamefont {E.}~\bibnamefont {Daw}}, \bibinfo {author}
  {\bibfnamefont {G.~D.}\ \bibnamefont {Hammond}}, \bibinfo {author}
  {\bibfnamefont {J.}~\bibnamefont {Hough}}, \bibinfo {author} {\bibfnamefont
  {R.}~\bibnamefont {Jones}}, \bibinfo {author} {\bibfnamefont {I.~W.}\
  \bibnamefont {Martin}}, \bibinfo {author} {\bibfnamefont {S.}~\bibnamefont
  {Rowan}}, \bibinfo {author} {\bibfnamefont {K.~A.}\ \bibnamefont {Strain}},\
  and\ \bibinfo {author} {\bibfnamefont {D.}~\bibnamefont {Williams}},\ }\href
  {https://doi.org/10.1088/1361-6382/abac42} {\bibfield  {journal} {\bibinfo
  {journal} {Class. Quantum Gravity}\ }\textbf {\bibinfo {volume} {37}},\
  \bibinfo {pages} {195019} (\bibinfo {year} {2020})}\BibitemShut {NoStop}%
\bibitem [{\citenamefont {Cagnoli}\ \emph
  {et~al.}(2000{\natexlab{a}})\citenamefont {Cagnoli}, \citenamefont {Hough},
  \citenamefont {DeBra}, \citenamefont {Fejer}, \citenamefont {Gustafson},
  \citenamefont {Rowan},\ and\ \citenamefont {Mitrofanov}}]{Cagnoli2000}%
  \BibitemOpen
  \bibfield  {author} {\bibinfo {author} {\bibfnamefont {G.}~\bibnamefont
  {Cagnoli}}, \bibinfo {author} {\bibfnamefont {J.}~\bibnamefont {Hough}},
  \bibinfo {author} {\bibfnamefont {D.}~\bibnamefont {DeBra}}, \bibinfo
  {author} {\bibfnamefont {M.}~\bibnamefont {Fejer}}, \bibinfo {author}
  {\bibfnamefont {E.}~\bibnamefont {Gustafson}}, \bibinfo {author}
  {\bibfnamefont {S.}~\bibnamefont {Rowan}},\ and\ \bibinfo {author}
  {\bibfnamefont {V.}~\bibnamefont {Mitrofanov}},\ }\href
  {https://doi.org/10.1016/S0375-9601(00)00411-4} {\bibfield  {journal}
  {\bibinfo  {journal} {Phys. Lett. A}\ }\textbf {\bibinfo {volume} {272}},\
  \bibinfo {pages} {39} (\bibinfo {year} {2000}{\natexlab{a}})}\BibitemShut
  {NoStop}%
\bibitem [{\citenamefont {Cagnoli}\ \emph
  {et~al.}(2000{\natexlab{b}})\citenamefont {Cagnoli}, \citenamefont
  {Gammaitoni}, \citenamefont {Kovalik}, \citenamefont {Marchesoni},\ and\
  \citenamefont {Punturo}}]{Cagnoli2000a}%
  \BibitemOpen
  \bibfield  {author} {\bibinfo {author} {\bibfnamefont {G.}~\bibnamefont
  {Cagnoli}}, \bibinfo {author} {\bibfnamefont {L.}~\bibnamefont {Gammaitoni}},
  \bibinfo {author} {\bibfnamefont {J.}~\bibnamefont {Kovalik}}, \bibinfo
  {author} {\bibfnamefont {F.}~\bibnamefont {Marchesoni}},\ and\ \bibinfo
  {author} {\bibfnamefont {M.}~\bibnamefont {Punturo}},\ }\href
  {https://doi.org/10.1063/1.1150607} {\bibfield  {journal} {\bibinfo
  {journal} {Rev. Sci. Instrum.}\ }\textbf {\bibinfo {volume} {71}},\ \bibinfo
  {pages} {2206} (\bibinfo {year} {2000}{\natexlab{b}})}\BibitemShut {NoStop}%
\bibitem [{\citenamefont {Massey}(1968)}]{Massey1968}%
  \BibitemOpen
  \bibfield  {author} {\bibinfo {author} {\bibfnamefont {G.}~\bibnamefont
  {Massey}},\ }\href {https://doi.org/10.1109/PROC.1968.6829} {\bibfield
  {journal} {\bibinfo  {journal} {Proc. IEEE}\ }\textbf {\bibinfo {volume}
  {56}},\ \bibinfo {pages} {2157} (\bibinfo {year} {1968})}\BibitemShut
  {NoStop}%
\bibitem [{\citenamefont {Speake}\ and\ \citenamefont
  {Aston}(2005)}]{Speake2005}%
  \BibitemOpen
  \bibfield  {author} {\bibinfo {author} {\bibfnamefont {C.~C.}\ \bibnamefont
  {Speake}}\ and\ \bibinfo {author} {\bibfnamefont {S.~M.}\ \bibnamefont
  {Aston}},\ }\href {https://doi.org/10.1088/0264-9381/22/10/019} {\bibfield
  {journal} {\bibinfo  {journal} {Class. Quantum Gravity}\ }\textbf {\bibinfo
  {volume} {22}},\ \bibinfo {pages} {S269} (\bibinfo {year}
  {2005})}\BibitemShut {NoStop}%
\bibitem [{\citenamefont {Cooper}\ \emph {et~al.}(2017)\citenamefont {Cooper},
  \citenamefont {Green}, \citenamefont {Collins}, \citenamefont {Hoyland},
  \citenamefont {Speake}, \citenamefont {Freise},\ and\ \citenamefont
  {Mow-Lowry}}]{Cooper2018}%
  \BibitemOpen
  \bibfield  {author} {\bibinfo {author} {\bibfnamefont {S.~J.}\ \bibnamefont
  {Cooper}}, \bibinfo {author} {\bibfnamefont {A.~C.}\ \bibnamefont {Green}},
  \bibinfo {author} {\bibfnamefont {C.}~\bibnamefont {Collins}}, \bibinfo
  {author} {\bibfnamefont {D.}~\bibnamefont {Hoyland}}, \bibinfo {author}
  {\bibfnamefont {C.~C.}\ \bibnamefont {Speake}}, \bibinfo {author}
  {\bibfnamefont {A.}~\bibnamefont {Freise}},\ and\ \bibinfo {author}
  {\bibfnamefont {C.~M.}\ \bibnamefont {Mow-Lowry}},\ }\href
  {https://doi.org/10.1088/1361-6382/aab2e9} {\bibfield  {journal} {\bibinfo
  {journal} {Class. Quantum Gravity}\ }\textbf {\bibinfo {volume} {35}},\
  \bibinfo {pages} {095007} (\bibinfo {year} {2017})},\ \Eprint
  {https://arxiv.org/abs/1710.05943} {arXiv:1710.05943} \BibitemShut {NoStop}%
\bibitem [{\citenamefont {Killbourn}\ \emph {et~al.}(1999)\citenamefont
  {Killbourn}, \citenamefont {Skeldon}, \citenamefont {Robertson},\ and\
  \citenamefont {Ward}}]{Killbourn1999}%
  \BibitemOpen
  \bibfield  {author} {\bibinfo {author} {\bibfnamefont {S.}~\bibnamefont
  {Killbourn}}, \bibinfo {author} {\bibfnamefont {K.}~\bibnamefont {Skeldon}},
  \bibinfo {author} {\bibfnamefont {D.}~\bibnamefont {Robertson}},\ and\
  \bibinfo {author} {\bibfnamefont {H.}~\bibnamefont {Ward}},\ }\href
  {https://doi.org/10.1016/S0375-9601(99)00596-4} {\bibfield  {journal}
  {\bibinfo  {journal} {Phys. Lett. A}\ }\textbf {\bibinfo {volume} {261}},\
  \bibinfo {pages} {240} (\bibinfo {year} {1999})}\BibitemShut {NoStop}%
\bibitem [{\citenamefont {Dmitriev}\ \emph {et~al.}(2009)\citenamefont
  {Dmitriev}, \citenamefont {Mescheriakov}, \citenamefont {Tokmakov},\ and\
  \citenamefont {Mitrofanov}}]{Dmitriev2009}%
  \BibitemOpen
  \bibfield  {author} {\bibinfo {author} {\bibfnamefont {A.~V.}\ \bibnamefont
  {Dmitriev}}, \bibinfo {author} {\bibfnamefont {S.~D.}\ \bibnamefont
  {Mescheriakov}}, \bibinfo {author} {\bibfnamefont {K.~V.}\ \bibnamefont
  {Tokmakov}},\ and\ \bibinfo {author} {\bibfnamefont {V.~P.}\ \bibnamefont
  {Mitrofanov}},\ }\href {https://doi.org/10.1088/0264-9381/27/2/025009}
  {\bibfield  {journal} {\bibinfo  {journal} {Class. Quantum Gravity}\ }\textbf
  {\bibinfo {volume} {27}},\ \bibinfo {pages} {025009} (\bibinfo {year}
  {2009})}\BibitemShut {NoStop}%
\end{thebibliography}%

\end{document}